\documentclass[pre,preprint,amsmath,amssymb,superscriptaddress,showpacs]{revtex4-2}
\usepackage{amsmath,amssymb,graphicx}
\usepackage{amsbsy}
\usepackage{latexsym}
\usepackage{color}
\usepackage{graphicx}
\usepackage{psfrag}
\usepackage[normalem]{ulem}
\usepackage{bm}
\usepackage{hyperref}

\begin{document}

\title{Some considerations about reviewing and open-access\\in scientific publishing}

\author{Paolo Politi}
\email{paolo.politi@cnr.it}
\affiliation{Istituto dei Sistemi Complessi, Consiglio Nazionale delle Ricerche, Via Madonna del Piano 10, 50019 Sesto Fiorentino, Italy}

\author{Giuseppe Gaeta}
\email{giuseppe.gaeta@unimi.it}
\affiliation{Dipartimento di Matematica, Universit\`a degli Studi di Milano \\ v. Saldini 50, 20133 Milano, Italy}

\author{Satya N. Majumdar}
\email{satya.majumdar@universite-paris-saclay.fr}
\affiliation{LPTMS, CNRS, Univ. Paris-Sud, Universit\'e Paris-Saclay, 91405 Orsay, France}

\author{Antonio Politi}
\email{a.politi@abdn.ac.uk}
\affiliation{Institute for Complex Systems and Mathematical Biology \& SUPA University of Aberdeen, Aberdeen AB24 3UE, United Kingdom}

\author{Stefano Ruffo}
\email{stefano.ruffo@sissa.it}
\affiliation{SISSA, Via Bonomea 265, I-34136 Trieste, Italy}

\begin{abstract}
Scientific research changed profoundly over the last 30 years, in all its aspects.
Scientific publishing has changed as well, mainly because of
the strong increased number of submitted papers and because of
the appearance of Open Access journals and publishers.
We propose some reflections on these issues.
\end{abstract}

\maketitle
\newpage

\begin{center}
\section{Preface}

\textit{Paolo Politi}
\end{center}

Scientific research changed profoundly over the last 30 years, in all its aspects:
the possibility to collaborate has been facilitated; 
the number of researchers has increased~\cite{USR};
computer has gone from being a calculation tool to be a conceptual tool~\cite{ilPonte};
bibliographic search was possible backward (who is cited in paper $X$?)
but extremely difficult forward~\cite{SCI} (who cites paper $X$?);
it was necessary going physically to a library to browse journals and
it was common to receive and to send postal cards to ask the author 
a paper copy of their article;
once an article was ready several months were necessary to disseminate it~\cite{arXiv}.

Scientific publishing has changed as well, mainly for two reasons:
firstly, the increased number of researchers has produced an increased number of
submitted papers without determining an equal increase of the capacity to process
such papers; secondly, the appearance of Open Access (OA) journals and publishers~\cite{OA} 
on the market. These journals ask authors to pay while keeping free access
to the journal content. A third novelty, only apparently a technical one, has 
played a major role: the introduction of metrics to evaluate 
researchers ($h$-index~\cite{hindex}) and journals (Impact Factor, IF~\cite{IF}).

It goes without saying that these changes have a strong impact on
the work of all researchers but a public discussion around them is lacking.
This small collection of short contributions has an obvious purpose but it
would like to have a second one.
The obvious purpose is to offer some basic information and some
considerations because the reader starts to think more deeply about these changes.
Hopefully the second purpose might be that other contributions
follow and that some proposals emerge to try to direct rather than only undergo changes.

We offer five contributions. 
The first contribution discusses why and how a ``low cost" journal was recently created.
It is written by Giuseppe Gaeta (University of Milan), who is in the editorial board of 
Open Communications in Nonlinear Mathematical Physics.
The following two contributions concern the editorial and review process
of articles submitted for publication and they are written by Satya N. Majumdar
(Université Paris-Saclay) and Antonio Politi (University of Aberdeen). 
SNM has served in the Editorial board of several journals
including Physical Review Letters, J. Phys. A: Math. Theo. (JPA), 
J. of Stat. Mech. (JSTAT), and also J. of Stat. Phys. (JSP). 
AP is serving as editor of Physical Review E since more than two decades; 
he has been twice part of the editorial board of J. Phys. A, and once of the European Journal of Physics D.

The last two contributions concern the Open Access process and are written by Stefano Ruffo
(SISSA Trieste) and by the author of this preface (Paolo Politi, CNR Florence).
SR is the SISSA Director and the
Chair of the Library Commission of CRUI (Conferenza dei Rettori delle Università Italiane). 
He offers several technical information to uderstand where we are and where we are going.
I am a board member of the Italian Society of Statistical Physics (SIFS) and in my contribution I 
argue against the ``pay to publish" model.

\newpage
   
\begin{center}
\section{On the cost of publishing, and a low cost alternative}

\textit{Giuseppe Gaeta}
\end{center}

There is no doubt that traditional publication of scientific journals involves substantial costs, and this is at the basis of the increasing costs which scientific institutions are required to cover for the publication of papers.

As it often happens, however, a closer look shows that -- albeit it is true that substantial costs are involved -- these are not always where one would expect them; this is also due to technological advances. For example, typographical composition of scientific texts, which once upon a time was a major task, and required skilled typographers, is by now carried on by the authors themselves, which in most cases provide \LaTeX files reproducing exactly the Journal format.

Similarly, communication with Authors and Referees is by e-mail, so at zero cost. As for refereeing, I feel very old when I remember that there was a time when you could get a print copy of the paper you had refereed, or maybe of the whole Journal Issue in which it appeared (I still have some on my bookshelf, and I could not avoid looking at them to be sure it was true); everybody knows how it works nowadays. Younger colleagues may be astonished in learning that some decades ago you could actually be \emph{paid} for writing an article, albeit not in all Journals: I remember the check when I wrote my first {\it Physics Reports}.

One would like to add among costs also the remuneration of Editors; I hope this is the case at least for Chief Editors carrying the responsibility of major Journals, but we know that in general the only advantage of sitting in an Editorial Board is to have some more visibility when you are young, and to please some friends when you get older and your name can somehow help the Journal (or your friends let you think this to be the case). Honors and Glory, so to say, but neither gold nor silver.

But, as I said, there are some \emph{real} costs. First of all physical printing, albeit much cheaper nowadays (just print from electronic files, and just the number of copies you actually need -- if need arises, other copies can be printed at any time, and this also eliminates stocking costs), has some cost. An even greater cost is related to (physical) distribution; this is usually taken care of by specialized agencies, but surely not for free. And then there is the cost of all the administration which is behind a publishing house: this means secretaries, administrative staff, and includes nowadays software like the (more and less loved) ``editorial systems'' which are used to communicate with Authors and Referees, exchange reports and so on.

And, as most publishers are commercial ones, the profit which should be rightly made out of a business, and which nobody wishes to deny.

It might be interesting to make some numbers to realize how costly it is to publish a scientific Journal, and what is the profit it may give, comparing this with other kinds of endeavor in the publishing business (I happen to have friends who run some small scientific publishing house, so I have some idea about this). But this is of no interest to us. The problem is that the total cost, production costs plus profit, must be covered somehow. And there are two sources: readers and writers. That is subscriptions, which may also take the form of access fees for single items, and ``contributions'' by Authors --  or in practice by their Institutions.

The balance is shifting towards the second option, as we all know (and as may seem odd to the general public, used to think you buy a book and some part of the money goes to those who wrote it); and this creates many problems, not only financial, which are discussed in a number of sources; including other articles in this collection.

One of the problems is that \emph{it costs a lot} to publish, especially in a good Journal, and especially if you produce a number of papers. Obviously, ``a lot'' is not a precise definition, and it should be compared with the funds available for this purpose (or in general) to the different groups. In come cases there are plenty of such funds, in other cases the national University system reaches some agreement with publishers to cover these costs (which means some  amount of taxpayers' money is flowing to the publishers); in other cases, simply there are no such money, and a number of Journals are \emph{de facto} forbidden.
Not so bad for certain ``pay-per-publish'' Journals; less good when it applies to serious and relevant Journals, where we are used to think merit and quality should be the criteria (and the reason why our libraries pay a subscription).

What I want to briefly discuss here is if there is a way out of this increasingly costly scheme. And I want to discuss this based on my own experience, or better on the experience of a specific Journal I have been associated with for several years, and which underwent a substantial change -- and actually a rebirth in a different skin -- after 27 years of life.

The new version of the Journal (let me make some advertising) is called 
\href{https://ocnmp.episciences.org}{Open Communications}
\href{https://ocnmp.episciences.org}{in Nonlinear Mathematical Physics}; 
the story of its ancestor is illustrated on its home page in an article by the Chief Editor, and here it will suffice to say that the Board of Editors (which were the editors of the other Journal as well) was not satisfied with the perspective of moving to an ``open access'' format which would require mandatory Article Processing Charge (APC), contrary to what had always been the philosophy of this group of editors.

The new Journal is completely free for both Authors and Readers, and it will be so permanently (thus not just for an initial period). How is it possible?

The method lies in reading carefully the list of costs which I gave before, and dispense with them. For example, the Journal is in electronic form only; and this eliminates the costs for printing and for (physical) distribution. There is no secretarial cost related to the running of the Journal, as the Journal is run directly by the editors: that is, rather than giving to a secretary the address of a colleague to ask him/her to referee a paper through an editorial system, I put the address myself in the system -- or better I write a real letter, not a form, asking him/her for the refereeing.

It should be said that this is possible because the numbers involved are relatively limited. The ``ancestor Journal'' published four Issues a year, with about ten papers per issue; needless to say, many more papers were submitted and then rejected (in many cases directly, either for evident lack of quality or for lying out of the scope of the Journal, i.e. without going to refereeing), but still things can be run in a simple way. Admittedly, it would be difficult to apply such a model on the scale of, say, the \emph{Physical Review}.

As for the cost of paying Editors, Referees and Authors... well, this problem has been solved earlier on by traditional publishers, as I recalled above. And there are no administration costs, as money is completely absent.

There is of course a catch: \emph{some} administration is needed: e.g. you should take care that the Journal is indexed in usual data bases, and also that an ISSN number is given, copyright forms should be collected and other legal matters should be taken care of. And these are things a scientist is usually not happy to deal with.

In our case, we were lucky that some specific foundations exist, taking care of allowing non-commercial scientific publishing, and we are hosted by one of these, i.e. the \href{https://www.episciences.org}{epi-sciences} foundation. This is not the only viable option: for example, several Universities have a University Press, and in many cases (e.g. for my own University) it is possible to publish with them at essentially the same conditions and with the same freedom, lack of costs, and covering of the ``boring'' tasks.

We are aware we are not the first to take such a step, and indeed we remembered that (in the words of Steve Jobs) ``you should not be ashamed of copying a good idea when you see one''. In some cases, commercial publishers turned out to be also willing -- maybe responding to the pressure of a strongly minded Editorial Board -- to have a diamond open access Journal, even in \href{https://www.aimspress.com/journal/MinE}{quite applied fields} (and personally I like to publish in such Journals).
There is of course a problem, even assuming no hostile action is taken against this model of publishing. New Journals take some time to establish their status, and in this time publishing in them may be not so good in view of different ``evaluation exercises'' run by national University or Research systems. This is specially relevant for young authors, but it affects also senior ones, if not personally at least in that their Departments may also be evaluated depending on where papers are published. In this sense it is essential that the databases used for such evaluation exercises include -- and do not hamper -- the ``free press'' Journals; this may be a problem, at least in principles, as some of the most used ones are owned by commercial publishing houses. I am not aware at the moment of any action in this sense by these databases; but in perspective it may be more safe to be sure evaluation is not based on tools provided by the same commercial firms whose products should be evaluated.

I would like to stress that I do not have anything against commercial publishers \emph{per se}. I would roughly classify traditional publishers in four categories: Learned Societies (e.g. AIP, APS, AMS, IOP); University Presses (e.g. Cambridge, Oxford, Princeton); serious commercial publishers (too many to list); and ``pathological'' commercial publishers (everybody will have his/her own example at hand or in the mailbox). I happily work with publishers of all the first three types. Actually learned societies and university presses which grow beyond a certain level often acquire the \emph{modus operandi} of serious commercial publishers. In any case, there is nothing wrong in publishing for a profit; it is also quite natural that commercial publishers try to maximize their profit. It is less natural, in my view, that scientific institutions do not always try to limit their expenses towards publication, and seem to accept whatever is proposed by the publishing industry; even more so when they administer public (that is, taxpayers') money.

I would like to stress the problem is not only an economical one, albeit I chose to focus on this issue: once the income comes from those who write, and not from those who read, it is possible (and it sometimes happens already) that groups with abundant funds do not produce high class science, but see however their work published. In the traditional scheme, a decline in the quality of published papers entails a decline in the number of subscriptions, hence a loss of income; but in this scheme there are no subscriptions, and there is a risk that Journals which were once of good quality will decline; or even worse that good quality Journals will reject good work due only to financial reasons.

Clearly, there are several ways to counteract these risks, first of all the integrity of publishers and editorial boards; and for major Journals this is not only to be expected but the only way to proceed.

But it may be good to know that, in particular for smaller-scale Journals, alternative routes do by now exist. If they will be viable depends first of all by the support by the community, which in this case is not financial but in terms of a steady flow of good quality work.

\newpage
\begin{center}
\section{Some reflections on the\\Publication/Review/Editorial process}

\textit{Satya N. Majumdar}
\end{center}

\subsection{Premise}

I have been an active researcher in Statistical Physics for the past $30$
years. 
The reflections/thoughts below are based
on my experience as an author, a reviewer and also as an Editorial board
member of the journals mentioned above.

Let me start with a basic problem that the academic world is facing 
today concerning the `cost of knowledge'. Several
private publishing houses with their brand name high profile
journals demand enormous article processing charges, in addition
to the already large subscription fees paid by the Universities and the 
research Institutions. I am not talking about non-profit/open-access 
journals like APS, JSTAT, SciPost that are academically
controlled and provide much better models. We scientists
do the actual research, also provide mostly
free services by reviewing the papers for such journals and
these publishing houses make enormous profits out of nothing really.
They use the scientists and we let them use us! This is
totally absurd and it has to stop! These private companies
need to scale down substantially their publication charges, subscription 
fees and in addition, they need to pay the referees substantially
for their services [see my comments later].

\subsection{Practical problems}
Having mentioned this basic problem, let me now turn to
some practical problems associated with review process and editorial
management. 

\subsubsection{Too many papers} 
I think everyone would agree 
that one of the major problems we face today is:  
there are far too many papers these days compared to the number of 
available referees who are not only competent, but are also willing to 
devote quality time for a serious review. When the editor contacts a 
referee, a majority of the referees do not bother to even reply. Some 
do, but they are typically busy and decline to review. They suggest some 
other names, who often reply in a similar 
way. Finally, some referees do an excellent job with useful and 
constructive suggestions to improve the paper, but their numbers are 
very few unfortunately. Quite often, the referee may agree to review but 
never gets down to the paper till she/he gets few reminders, and finally 
just a cursory glance (10 minutes typically!) decides the 
fate of the paper. The outcome is a `generic' report with very little 
values. It is actually not very difficult for the authors to figure out 
quite quickly if the referee has read the paper or not. All of us, as 
authors, have unfortunately faced this situation quite frequently.

\subsubsection{Referee selection process} 
In some journals like 
J. Stat. Mech. or J. Stat. Phys., it is an Editorial board member 
(typically from the field of the paper) who selects the referees.
The outcome is invariably better because the board member is a
practicing scientist and usually can decide very quickly who could
be the potential and appropriate referees. In contrast, in journals
such as PRL, a desk editor (some of them are of course extremely competent)
selects the referee (typically from keywords and the available referee
data bases). Since the desk editor is not a practicing scientist, even
though they may be very competent, there is always the possibility
of a judgemental error in selecting the referee. 
As a result, often the paper lands up with a referee who
is hardly interested in the subject, or doesn't have a good
idea of the main issues in the field. Consequently, one often
gets a `generic' and rather `shallow' report with comments like 
`good work, but not interesting enough for a broader audience'.
Such reports are absolutely useless and some referees latch on to
one of those vague criteria like `broad importance' etc. to quickly
reject a paper without even trying to read and grasp the content of the
paper. Again, I am sure that many of us, as authors, have invariably
faced this problem far too often.

\subsubsection{Lack of incentives for referees} 
One of the major problems
today is that there are not much incentives for a referee to do the
review, unless the paper is of direct interest to their own work.
Hence the quality of the scientific reviews has really taken
a huge plunge compared to, even, $20$ years back. In my opinion,
the review process has to undergo some major reforms (see my comment further
down).

\subsubsection{Too many journals competing for higher impact factors} 
Unfortunately the scientific community as a whole has become too 
sensitive to metrics such as impact factors, citations, $h$-indices etc. 
This is a direct result of big-grant driven research. People want to 
publish their results in high profile journals as quickly as possible, 
but often spare little concerns on the quality of the work
and the presentation. To 
improve the impact factor, the journals often encourage reviews (they 
typically get more citations) or special issues. I remember from my days 
as a student that we learnt a lot by reading lecture notes, e.g.,
the Les Houches lecture notes. 
But these days, very few people bother to write really pedagogical 
lecture notes, since typically they take time and
often may not fetch much citations! So, 
this is general problem that the whole academic community is facing, and 
I think one needs serious debates about the quality of scientific papers, 
and not the quantity or the number of citations it brings. A better
(hypothetical) criteria for the quality of a paper should be whether (i) 
the results will remain valid say, even after 50 years and (ii) whether people
will still find the paper useful after 50 years. Most of the
papers published today won't satisfy this criteria.

\subsubsection{Rise of predatory journals} 
Each day most of us receive
many emails from all types of journals asking us to become a member
of its Editorial board or submit a paper. They are extremely
annoying to say the least. Most of us ignore such mails,
but some take them seriously and these journals survive by those
judgemental errors on the part of some researchers. 
Such predatory journals, causing nothing but pure nuisances,
should be aggressively stopped.

\subsection{What's the way out?} 

It is, of course, not easy to come up 
with quick remedies to all the issues and problems I mentioned above.
Here are some suggestions for the way out of some of these problems.

\subsubsection{Review process}
I think, of all the problems, the most serious one concerning the 
review process that requires immediate attention is how to provide some 
incentives for referees to do their job seriously. For a very long 
time, referees have been doing their work voluntarily. But the time has 
changed and it requires new initiatives. Personally, I think one simple 
solution is to provide substantial financial incentives to referees 
(and not just some token fee). As I mentioned earlier, several private 
publishing houses are making enormous profits out of publishing our 
articles. In my opinion, if we want to improve the quality of the 
refereeing process, the referees should be paid a substantial fee for 
their services, especially from these profit making journals.

\subsubsection{Journals}
We need to encourage our students and colleagues to publish
more in non-profit/open-access journals such as JSTAT, SciPost etc.
which are excellent models about how the publication process
should be conducted. The academic community has to regain the control
of the publication process, and not leave it to private buisnesses.
For example, when senior scientists sit in evaluation committees (for
promotions, grants etc.), they need to valorise the quality and the
content of a paper, and not just the impact factor of the journal where
it is published or the number of citations it has received. Again
time is an issue: most people want a quick number (like h-index)
or the number of Nature/Science/PRL 's to evaulate a person.
But if we do not de-brand these journals or the metrics ourselves, 
we will pay for it dearly in the long run (and are already paying for it!).

I wish to thank E. Trizac for his valuable comments and inputs
on this brief article.

\newpage
\begin{center} 
\section{What is happening to the reviewing process\\of publications in scientific journals?}

\textit{Antonio Politi} \end{center}

It is not clear to me which problems are specific to the Stat Mech area. I am going to argue in general terms.

\subsection{Premise}

The major problem resides outside the reviewing/publication process: it is the far too large number of preprints continuously produced: it is not possible to judge them in a reasonable way, no matter which method is going to be used. I am deeply convinced that one should seriously address this problem.

The number of submissions has greatly increased because of the appearance on the international arena of India and China. On average, the quality of submissions coming from those areas is low but it has increased: at the moment, it makes the reviewing process very difficult because it is not obvious to recognize immediately poor papers.
I cannot say whether the number of submissions per single person has increased in the last years: this happened many years ago with the new automatic technologies, but I suspect
that now is almost stationary.

I am definitely convinced that many submissions are of low quality, in a way that it is not
obvious how to spot, since I have the simultaneous feeling of the low quality of researchers
which are the potential referees: this is a vicious cycle.
How large should the referee pool be? Too large means exactly that poor researchers accept poor papers.

\subsection{Issues}

\subsubsection{Open access}

Bad idea, in so far as the payment is the authors’ burden: it discriminates those who obtain valuable results without having been funded. In a normal world, such authors should be rewarded for the low cost of their research!
If societies and funding agencies cover upfront expenses (the so-called diamond open access) the objection dies down.
I am aware that the Max Planck Society in many cases cover such expenses upfront. Can one extend this strategy to all countries and disciplines? Very unlikely.

Much more feasible is the idea of pushing for green open access: this is a point where physics, math and a few other disciplines could distinguish themselves from the rest of the scientific world, by introducing a mandatory archival of the accepted version (APS already accepts this policy). Being afraid of changes during the proofreading process is totally idiotic: a serious journal goes back to the referees if relevant changes are spotted.

\subsubsection{Reviewing process}

Is it really necessary to undergo a reviewing process?

This idea is often proposed, but I am afraid that in the current world it clashes dramatically with what is becoming the main motivation to publish: validating the use of funds (fellowship, research project, travel funds, own salary). Can one disentangle publications produced to confirm a fair use of funds from a substantial scientific progress (reducing the number of $``$checks$"$ ?)  positive answer to this question would greatly contribute to solve the problem mentioned in the premise.

\subsubsection{Referees}

The willingness of our fellow colleagues to contribute to the reviewing process is very heterogeneous: some may have good reasons, but every senior scientist should be encouraged to contribute, as part of their duties. Unfortunately, this is not the case and it is difficult to coordinate the activity spread over different journals.

Paying referees is a very bad idea, because of the resulting conflicts of interests.

Recognition of $``$outstanding referee$"$  does not help much in the long term.

Discount on publication charges (if charges have to be applied at all)?

\subsubsection{Publishers}

Journals run by scientific societies (e.g. APS and IOP) were used to provide a good service to the community. My personal experience is that journals run by societies were functioning in a reasonably good way.

In the last years I have witnessed an increasing tendency to compete (especially because of the emergence and growth of private publishers) to maintain and increase the readership.

Altogether, scientific publishers kept aggregating in fewer groups thereby gaining increasing weight (Nature-Springer above all). Simultaneously, journals proliferated and keep doing so under the hope of attracting more papers.

This method works because journal subscriptions are typically arranged in large portfolios, rather than handled separately.

\subsection{Suggestions}

Given the increased concentration of publishers, the only way to react is via societies, which can have enough power to negotiate agreements.

Funding agencies have also a crucial power: can we convince them to contribute, without linking the funds to pre-funded projects?

\newpage

\begin{center}
\section{Open access and the strategy\\of transformative agreements}
\label{sec.SR}

\textit{Stefano Ruffo}
\end{center}

At the beginning of the 90’s, with the development of the Internet, it became clear that the circulation of scientific information through “preprints" sent by snail mail could cease and that it could take place using the web. The most successful project was that of Paul Ginsparg in Los Alamos (USA) in 1992, which has now become the arXiv of Cornell University. SISSA also launched a project that had a minor impact.

The first electronic journals began to appear in those years, and in 1997 SISSA launched the Journal of High Energy Physics (JHEP), a journal “made by scientists for scientists", which has now become the world's leading journal in the field of high energy physics. The JHEP model was replicated in the following years with journals of similar impact: JCAP, JSTAT, JINST.

It was then important to demonstrate that innovative and competitive journals can be made outside the perimeter of major publishers (Elsevier, Nature, Springer, Wiley,…).

It was equally important to identify “business models" that would ensure the long-term sustainability of the journals. It is clear that publishing cannot have a zero cost, the non-eliminable costs are those of the editorial management of the journal and of the “peer review". So the questions that must be asked are: 
i) what is the "right" cost, ii) who should be remunerated, iii) should the research institution pay to allow reading with a subscription model, “pay to read ", or pay to allow the author to publish in open access, “pay to publish ".

Publishers can be private, public, semi-public, they can aim for profit if private or for the reinvestment of budget profits if non-profit. In short, the panorama is very wide and varied, it is an “ecosystem" in which new “species" continually appear and become extinct.

In the early 2000’s, some “principles" broke into this “ecosystem" with the declarations of Budapest (2002), Bethesda (2003), Berlin (2003). Among these, the principle with the greatest impact is that related to the right “for all” to read at no cost, the so-called “open access”. The ethical basis for this principle is the simple fact that, if research is publicly funded, the results must be publicly accessible.

The advantages resulting from the implementation of this principle are also evident: free access to scientific literature makes the knowledge contained in the scientific literature available to all researchers and facilitates its advancement thanks to the contribution of everybody. In addition, “open access" can allow a more effective knowledge transfer and facilitate industrial innovation. The libertarian ideological element of this principle is also non negligible, it is perhaps what has most moved the minds and the will of its supporters.

On the wings of this principle, a new business model emerged in the 2000’s which envisages that “open access" is supported by the author, through funding from the institution, public or private, for which he works. Subscription costs are completely removed and everyone has free online access to all publications. The most striking success story of this business model was that of PloS ONE, a journal published by the Public Library of Science, founded in 2006. The aim of this initiative was to create “open access" at a cost slightly above production costs, making the profit null or negligible. The hope that this model could guarantee “fair” publication prices, within everyone's reach, has faded away from year to year.

Meanwhile, the “open access" movement found an ever-increasing number of supporters and began to have, at the beginning of the 2010’s, important institutional support at the European level. In 2012 the “Finch Report” was published in the UK in which a decisive transition to the “open access” publication of the research results was urged through the creation of specific funds. In 2013, the "Science Europe" Association was founded, which brought together many European academies and research institutions in favor of “open access" (for Italy only INFN joined). In those years, the European Commission established that all research financed with European funds should be “open access".

Since few journals were ready for a transition to “pay to publish", the phenomenon of the “hybrid" began to emerge: in addition to the subscription costs, the author was also asked for a contribution to make the publication “open access”, the so-called "Article Processing Charge" (APC). This phenomenon, superimposed on the growing cost of journal subscriptions, produced a further increase in publication costs by universities and research institutions, causing the unsustainability of libraries' budgets, which, unable to bear the costs anymore, began to cut subscriptions.

And the researchers, where did we leave them? Physicists and mathematicians, in the excitement of the early 1990’s over the availability of web resources, have continued, and indeed greatly increased, the deposit of their articles in the arXiv, which has grown in size and reputation. Thanks to much more substantial financial resources ("grants"), researchers in the field of life sciences have absorbed the increased costs of publication without major problems and have been able to afford the new standards of "open access".

The SISSA journals JHEP and JCAP, after an initial agreement with the Institute of Physics (IOP) and then with Springer, found a way to finance “open access" with the creation of the SCOAP3 Consortium, which involves a network of about three thousand libraries and of CERN. Funds come from CERN and libraries, and the publishers involved are Springer, IOP, Elsevier, Oxford University Press, Hindawi, Jagellonian University with a dozen of journals in the field of high energy physics.

In 2015 an important "White Paper" of the Max Planck Digital Library (MPDL) was published which concludes: "All indications are that the money already invested in the research publishing system is sufficient to enable a transformation that will be sustainable for the future". It is the trigger for the process of the so-called "transformative contracts" between institutions and publishers. The argument is simple: from the data it appears that researchers are paying on average for each paper, within the subscription business model,  about 4000 Euros, a high cost for a “good" that has restricted access and usability. In addition, the phenomenon of “double dipping" has spread so much that in order to read a journal we do not only pay the cost of the subscription, but also the APC’s. An estimate of how much it would be “fair" to pay for each article published in "open access" is made in the document at about 1500-2000 Euros on average. The Open Access 2020 initiative (OA2020), launched by MPDL, aims at guiding the international process of “transformation" of all subscription based contracts into "open access" in which only APC’s are paid, aiming at the average cost per paper mentioned above. It is a “transformative process" whose goal is a "transition" from the “subscription" (S) model to that of “open access" (OA). Ulrich Poeschl, chemist at the Max Planck Institute in Mainz and one of the main promoters of this initiative, sees this process as a “chemical reaction" (S → OA), in which an “activation energy" is needed which could also lead temporarily to higher costs, but moderately and only in the transformation phase, until the system would settle into the “OA state” which would have lower costs.
We are in the midst of this “transformation", which is showing considerable success in the recent months also in Italy, where agreements have been signed with Springer, Wiley, American Chemical Society, Cambridge University Press, Emerald, de Gruyter and negotiations are underway in 2021 with a dozen of other publishers. The Italian model is based on agreements that universities and research centers have made with the Conference of Italian University Rectors (CRUI) which entrusts to the CRUI itself the conduct of negotiations with publishers. The Group for the Coordination for Access to Electronic Resources (CARE) operates within the CRUI, it is a team that has developed considerable skill in conducting negotiations over the years and which has recently obtained important results.

I was invited to participate in the recent Open Access Summit, organized by OA2020, which was held remotely on 14 December 2020. During the meeting, the most recent data on the “transformative process" were presented. The Max Planck Society now publishes over 80\% of its articles in “open access" with 20\% of the most important publishers. The S → OA transition is showing a strong acceleration centered in the period 2018-2019 and many countries now cover with “transformative contracts” from 30\% to 50\% of publications: Sweden, The Netherlands, Norway, Hungary, Austria, Finland, Switzerland, Germany, UK, Ireland. Italy is just at the beginning of this process. The OA2020 consortium looks forward to the "XV Berlin Open Access Conference" which will be held before the summer.

\newpage
\begin{center}
\section{Against the ``pay to publish" model}

\textit{Paolo Politi}
\end{center}

Two justifications are given to drop out the ``pay to read" model in favour of the ``pay to publish":
(i)~The results of publicly funded research must be immediately available and free of charge to the taxpayer who funded that research;
(ii)~Subscription prices have risen much more than average inflation (and the bulk of the work is provided by the academic community, which is unpaid).
The first consideration is welcome and the second one is true but both should be considered more carefully.

First of all, since the vast majority of researchers are public employees, 
certain obligations should exist regardless of specific funding and I don't dare to think what would happen 
by putting together researchers from different countries, with different funding and different rules, in the same article. 
Furthermore, should public funding for research really require \textit{universal and immediate access to results}? 
Since ``results" are not just publications, should researchers make public the raw results of experiments or simulations?
And what about the writing of monographs and textbooks? Should they also pass to a ``pay to publish" model? This is not a specious
question if you think that in Humanities the writing of monographs is the standard way to disseminate results and move up the career ladder.
I think we should consider that different disciplines have different ways to disseminate results and that public repositories
are the way to make results public, not Open Access (OA) journals (see below).

The second consideration  about subscription prices 
is certainly true but publishers offering OA journals either are the same publishers whose subscription costs
have increased or they are new predatory publishers. It is obvious that the problem of unsustainable subscription costs must be faced and that
it is necessary to face publishers at the supranational level. It is also necessary to encourage the creation of new journals.

I see various reasons to be against the ``pay to publish" strategy.
\begin{itemize}
\item The quality of many OA journals is unacceptably low and there has been an explosion of scam/junk journals that 
are polluting scientific publishing and are stalking the community. In some cases, with ``pay per publish" and
``pay per give a talk" you can buy your CV.

\item If the researcher pays to publish a pernicious short circuit is created between the authors and the journal, 
which have the same goal: that the article is published. 
When journals circulated only in paper there were ``physical" limits to the number of articles acceptable and publishable in a single issue, now this is no longer true.
The magazine must not have an interest in maximizing the number of published articles and 
the author must not be able to think that a good place to publish the paper will be found ayway  because (s)he has the money.

\item A practical example of this unhealthy relationship: the OA journal (even a ``serious" one) invites someone 
to be guest editor of a special issue on a certain topic, asserting the recipient would be the most suitable person 
and for this job the publisher offers the guest editor to publish for free. In practice this happens because the guest editor assures the journal 
a certain number of contributors who will pay instead!
And if the guest editor is good enough to convince a few authors (s)he will even get a free hard copy of the special issue.
But I'm sure, as we are at the livestock market, the guest editor can negotiate better terms and obtain
discounted fees or green cards to invite someone to publish for free.

\end{itemize}

Supporters of ``open access" state that publishing on such journals (gold OA) is not mandatory because authors can use OA archives (green OA)
if funding requires research results are made public.
But the choice between OA and not OA journals is less and less 
an equal choice because Impact Factor (IF) plays a major role and IF of an OA journal
is typically higher.  
The consequence would be that only people without money would use green OA 
and this would create a split in the scientific community (between different countries, between different research domains, 
between different research groups).

The IF inevitably leads to discuss the relation between publishing and research funding,
because the latter is more and more based on pre-funded projects.
If we want to get rid of the IF (do we?) it is necessary that the evaluator does not use it. 
Switching to OA journals will reinforce the importance of metrics, because they
introduce the ``dollar" factor: the higher the IF of a magazine, the more it can afford to charge publication costs. 
The higher these are, the smaller the circle of those who can try to publish in a high IF magazine. 
(Some journals charge authors just to start the editorial process.)

It is unrealistic to think that the majority of researchers affiliate to organizations that will be able to sign 
transformative agreements with Publishers (see Sec.~\ref{sec.SR} by SR) and 
it is unrealistic to think that researchers can afford to regularly pay to publish in OA journals.
Therefore, OA journals
will lead (are already leading) to a split in the scientific community, with a class A and a class B community: 
in the former class there is money to publish, there are high IF publications and there is access to funding; 
in the second class nothing or almost nothing of all this. 

The well founded requests to make public the results of public research and to counteract the soaring prices
of subscriptions could be met with
public repositories, with collective agreements with publishers, and with the birth of new journals,
always keeping in mind that different communities of scholars may have different ways to disseminate their results.
The guarantee of green Open Access while a part of the scientific community moves towards the gold Open Access
would not be the right solution.


\begin{thebibliography}{99}
\bibitem{hindex}
The $h$-index (after J. E. Hirsch) is a metric that measures both the productivity and citation impact of the publications of a scientist.
The $h$-index is defined as the maximum value of $h$ such that the given author has published $h$ papers that have each been cited at least $h$ times.
An italian physicist (Paolo Rossi, University of Pisa) has proposed a provocative extension of $h$-index for the evaluation and ranking
of medical doctors, as the (maximal) numbers of their patients who survived $h$ years to the cure.
I might propose the same for the evaluation of prosecutors: $h$-index is the number of defendants sentenced to at least $h$ years.
\bibitem{IF}
The Impact Factor of a journal is the yearly average number of citations of articles published in the last two years in that journal.
It is provided with three decimal digits while the distribution of citations is very broad.
\end{thebibliography}
\end{document}